\documentclass{article}

\usepackage{PRIMEarxiv}
\usepackage{amsthm}

\usepackage{makecell}

\usepackage[utf8]{inputenc} % allow utf-8 input
\usepackage[T1]{fontenc}    % use 8-bit T1 fonts
\usepackage{hyperref}       % hyperlinks
\usepackage{url}            % simple URL typesetting
\usepackage{booktabs}       % professional-quality tables
\usepackage{amsfonts}       % blackboard math symbols
\usepackage{nicefrac}       % compact symbols for 1/2, etc.
\usepackage{newunicodechar}
\newunicodechar{ }{\,} % handle U+202F NARROW NO-BREAK SPACE
\usepackage{microtype}      % microtypography
\usepackage{lipsum}
\usepackage{fancyhdr}       % header
\usepackage{graphicx}       % graphics
\graphicspath{{media/}}     % organize your images and other figures under media/ folder

\usepackage[numbers]{natbib}
\usepackage{amsmath}
\usepackage{subcaption}
\usepackage{algorithm}
\usepackage{algorithmic}
\usepackage{natbib}
\usepackage{amssymb}

\title{Robust Super-Capacity SRS Channel Inpainting via Diffusion Models}

\author{%
Usman~Akram\textsuperscript{\dag} \textsuperscript{\ddag}, 
Fan~Zhang\textsuperscript{\dag} \textsuperscript{*}, 
Yang~Li \textsuperscript{*}, 
and~Haris~Vikalo \textsuperscript{\ddag}\\[1ex]
\textsuperscript{\dag}These authors contributed equally to this work as joint first authors.\\[0.5ex]
\textsuperscript{\ddag} Department of Electrical and Computer Engineering, The University of Texas at Austin, TX, USA\\
\textsuperscript{*} Standards and Mobility Innovation Lab, Samsung Research America, Plano, TX, USA\\
E-mail: usman.akram@austin.utexas.edu, hvikalo@ece.utexas.edu, fan.zhang@samsung.com, yang.li1@samsung.com\\[0.5ex]
This work was conducted as part of the author’s summer internship at Samsung Research America.
}

\begin{document}
\maketitle

\begin{abstract}
Accurate channel state information (CSI) is essential for reliable multiuser MIMO operation. In 5G NR, reciprocity-based beamforming via uplink Sounding Reference Signals (SRS) face resource and coverage constraints, motivating sparse non-uniform SRS allocation. Prior masked-autoencoder (MAE) approaches improve coverage but overfit to training masks and degrade under unseen distortions (e.g., additional masking, interference, clipping, non-Gaussian noise). We propose a diffusion-based channel inpainting framework that integrates system-model knowledge at inference via a likelihood-gradient term, enabling a single trained model to adapt across mismatched conditions. On standardized CDL channels, the score-based diffusion variant consistently outperforms a UNet score-model baseline and the one-step MAE under distribution shift, with improvements up to 14 dB NMSE in challenging settings (e.g., Laplace noise, user interference), while retaining competitive accuracy under matched conditions. These results demonstrate that diffusion-guided inpainting is a robust and generalizable approach for super-capacity SRS design in 5G NR systems.
\end{abstract}

% keywords can be removed

\section{Introduction}

Accurate channel state information (CSI) is essential for reliable data recovery in massive MIMO and millimeter-wave systems, and has therefore attracted considerable attention in the research community. However, existing CSI acquisition methods exhibit significant limitations in practical settings, particularly under user mobility and channel model mismatch. Traditional channel estimation algorithms \cite{trad_1, trad_2}, rely on prior statistical knowledge or accurate parametric models, making them sensitive to model mismatch and high mobility. Compressive sensing–based approaches \cite{trad_3,trad_4,trad_5}, reduce pilot overhead but incur high computational complexity. Deep learning methods enable efficient inference once trained, but often lack robustness to deployment-time distribution shifts.

In 5G New Radio (NR), two uplink beamforming strategies are standardized: reciprocity-based beamforming, which uses sounding reference signals (SRS), and codebook-based beamforming. While SRS-based methods yield more accurate CSI, they face two key limitations:
\begin{enumerate}
\item \textbf{Low coverage}: As shown in \cite{Yang2023AiChAug}, SRS-based methods underperform precoding matrix indicator (PMI) approaches in low-SNR settings, limiting their effectiveness when the user equipment is far from base station.
\item \textbf{Limited SRS resource}: 5G NR standards define a fixed number of SRS allocation slots, constraining the number of simultaneously served users.
\end{enumerate}

Our prior work \cite{prior_inpainting} used a masked autoencoder with a Vision Transformer backbone for SRS-based channel inpainting, improving coverage and SRS efficiency. However, that approach relied on fixed-pattern reconstruction and degrades under mask mismatch, interference, or non-Gaussian noise. This paper introduces a generative diffusion-based framework that jointly denoises and inpaints while integrating system-model knowledge via a log-likelihood gradient at inference, enabling a single trained model to generalize across deployment-time distortions. To our knowledge, this is the first application of diffusion-based inpainting to SRS-based CSI acquisition.

%The remainder of the paper is organized as follows. Section II formulates the system model. Section III details the proposed diffusion-based inpainting framework. Section IV presents simulation results under various conditions, and Section V concludes with a discussion of key findings.

\section{System Model}

We consider uplink CSI acquisition in 5G NR via Sounding Reference Signals (SRS), where the user equipment (UE) transmits known Zadoff–Chu sequences \cite{holma2011lte} to the base station (gNB) for channel estimation. Let $N$ denote the total number of subcarriers and $A$ the number of receive antennas. A subset of subcarriers $\mathcal{S} \subseteq {1,\dots,N}$ is selected for SRS transmission. For each active subcarrier $k \in \mathcal{S}$ and antenna $a \in {1, \dots, A}$, the received signal is given by
\begin{align}
 Y[k,a]=H[k,a]X[k,a]+N[k,a],
\end{align}
where $X[k,a]$ is the known SRS symbol, $H[k,a]$ is the complex channel gain, and $N[k,a] \sim \mathcal{CN}(0, \sigma^2)$ denotes additive white Gaussian noise. A coarse channel estimate is obtained via zero-forcing, i.e., 
$\hat{H}[k,a]=Y[k,a]/X[k,a]$.
This yields a noisy and partial estimate $\hat{H}$ over the observed subcarrier–antenna pairs $(k,a) \in \mathcal{S} \times {1,\dots,A}$. Enhancing $\hat{H}$ is thus formulated as a channel denoising and inpainting problem, for which various deep learning models have been proposed. After refinement, the channel is interpolated to unobserved subcarriers $k \notin \mathcal{S}$. In typical SRS designs, the set $\mathcal{S}$ is selected uniformly across subcarriers, with the comb pattern determining the spacing. For instance, in a comb-4 pattern, every fourth subcarrier is allocated for SRS transmission. 

\begin{figure}[h]
    \centering
    \begin{subfigure}[b]{0.475\textwidth}
        \centering
        \includegraphics[width=\textwidth]{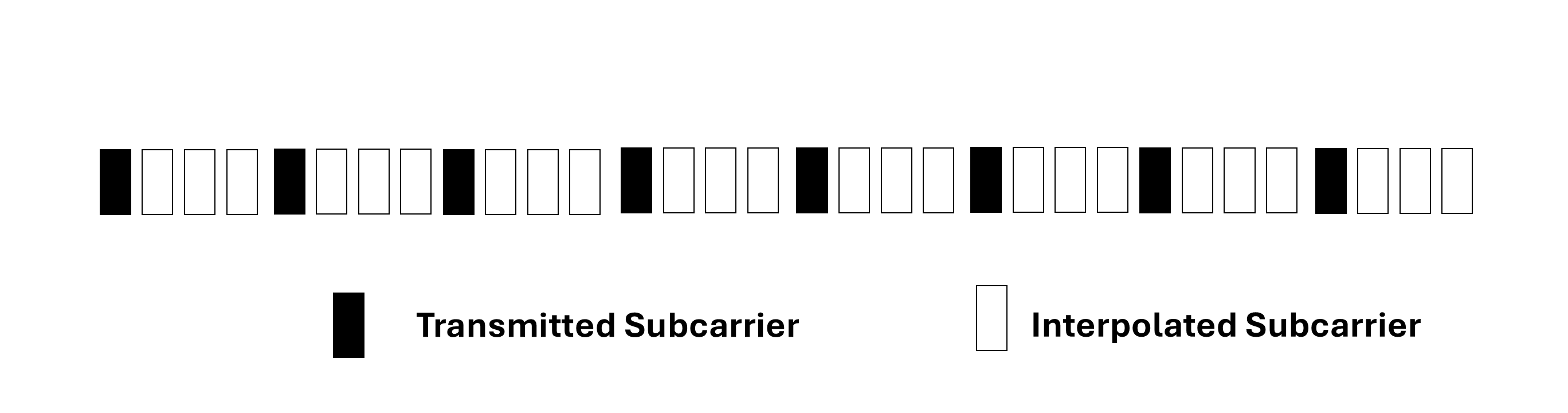}
        \caption{\small Standard comb-4 SRS allocation for $N=32$ subcarriers: a single user is allocated 8 uniformly spaced tones (every $4^{th}$ subcarrier) for uplink SRS transmission.}
        \label{fig:subfig1a}
    \end{subfigure}
    \hfill
    \begin{subfigure}[b]{0.475\textwidth}
        \centering
        \includegraphics[width=\textwidth]{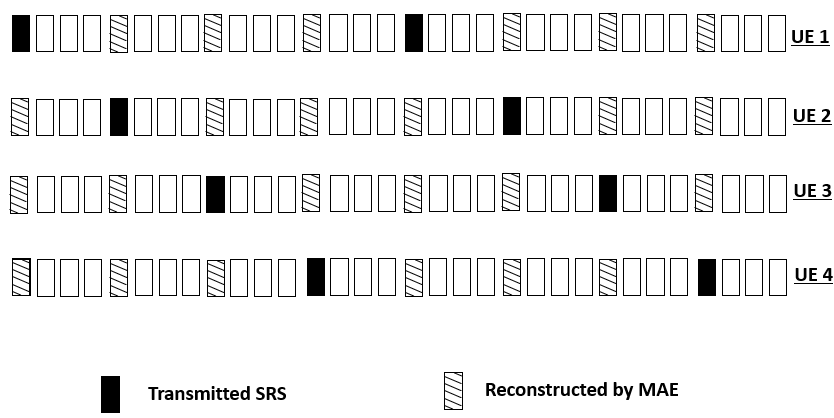}
        \caption{\small Masked SRS allocation for 4 users using the same 8 comb-4 subcarriers. Each user transmits on a distinct 25\% subset (2 subcarriers) and relies on reconstruction (e.g., via masked autoencoders) for the remaining 75\%.}
        \label{fig:subfig1b}
    \end{subfigure}
    \caption{SRS allocation under comb-4 pattern with $N=32$ subcarriers. (a) A single UE is assigned 8 uniformly spaced tones. (b) With sparse non-uniform masking and channel inpainting, the same resources support 4 users simultaneously.}
    \label{fig:fig1}
\end{figure}

Our prior work \cite{prior_inpainting} explored sparse, non-uniform SRS allocation within the constraints of 5G NR, as illustrated in Fig.~\ref{fig:subfig1a}–\ref{fig:subfig1b}, with two main objectives:
\begin{enumerate}
\item Increase SRS resource capacity to support more users.
\item Improve SRS coverage by concentrating power on a smaller subset of subcarriers.
\end{enumerate}
This approach demonstrated improved spectral efficiency and adaptability, particularly under tight SRS constraints.

However, the masked autoencoder framework used in \cite{prior_inpainting} exhibits limited robustness. In particular, its performance degrades under distribution shifts (e.g., novel masking patterns, additional noise, or user interference) due to overfitting to a limited set of training conditions. Although the framework accommodates flexible, non-orthogonal SRS masks, the combinatorial growth in masking configurations makes generalization difficult for purely supervised models. In this work, we propose a diffusion-based extension that improves robustness by incorporating domain knowledge at inference and mitigating overfitting under mask mismatch and data scarcity.

% \begin{figure}[h]
%     \centering
%     \begin{subfigure}
%         \centering
%         \includegraphics[width=0.4\textwidth]{Fig_1_a.png}
%         \caption{Unmasked SRS allocation for $N=325$ and $N_{\text{comb}}=45$}
%         \label{fig:subfig1a}
%     \end{subfigure}
%     \hfill
%     \begin{subfigure}
%         \centering
%         \includegraphics[width=0.4\textwidth]{Fig_1_b_modded.png}
%         \caption{75\% Masked SRS allocation for $N=325$ and $N_{\text{comb}}=45$ which allows us to multiplex 4 users}
%         \label{fig:subfig1b}
%     \end{subfigure}
% \end{figure}

\vspace{0.2in}

\vspace{2.5em}
\section{Proposed Method}

\subsection{Algorithm Design}

We introduce a diffusion-based framework for CSI inpainting that decouples training from deployment-time conditions, enabling a single model to generalize across a wide range of SRS masking patterns, noise levels, and interference regimes. Unlike prior masked autoencoder approaches, trained and evaluated under fixed supervision, our method leverages likelihood-guided inference to adaptively refine partial CSI estimates at test time. This formulation not only improves robustness under distributional shift but also enables principled integration of system knowledge, such as physical noise models or masking constraints, through the inference objective itself.

A seemingly natural strategy is to train a standard diffusion model for denoising and then perform channel inpainting at inference using methods like Noise-Conditioned Score Networks (NCSN) \cite{MIMO_Score_Est} or Diffusion Posterior Sampling (DPS) \cite{DPS}. However, we find that this approach yields subpar performance in the non-uniform SRS masking setting, where the missing data structure is highly nontrivial. This motivates a more targeted training scheme that integrates inpainting directly into the denoising process.

Instead, we adopt a design inspired by ambient diffusion \cite{ADPS, ADPS2}, which jointly learns denoising and inverse recovery when clean inputs are unavailable. While ambient diffusion is typically applied to linearly corrupted observations without access to ground truth, our setting is different: we do have clean training examples, and the corruption arises from structured masking. We therefore adapt the ambient diffusion framework by directly training on masked inputs to simultaneously perform denoising and inpainting. This approach aligns training with inference-time objectives and yields models that are more robust to varying masking patterns, noise levels, and other deployment-time degradations.

We explore two diffusion-based formulations for channel inpainting: score-based models using variance-exploding stochastic differential equations (VE-SDEs), and denoising diffusion probabilistic models (DDPMs) based on variance-preserving SDEs (VP-SDEs). Although the ambient diffusion framework was originally developed for DDPM-style generation, its core principle of joint denoising and inverse recovery from corrupted inputs naturally extends to the VE-SDE setting. As demonstrated in Section~\ref{sec:experiments}, the VE-SDE variant consistently outperforms its VP-SDE counterpart across a range of degradation scenarios, highlighting its robustness for non-uniform SRS inpainting.

For score-based diffusion, we adopt a noise-annealing strategy in which the variance levels $\{\sigma_{z_i}^2\}_{i=1}^L$ follow a geometric progression across $L$ steps. Let $f_\theta$ denote the score model with parameters $\theta$, and let $\mathcal{A} \sim p(\mathcal{A}_{\text{train}})$ be a sampled binary mask. For a clean channel matrix $H$, we sample an index $i \sim \mathcal{U}[1, L]$ and draw noise $z \sim \mathcal{N}(0, \sigma_{z_i}^2)$ to construct the corrupted input
$\tilde{H}_i=\mathcal{A}\odot H + z$. The model is then trained to minimize the following denoising score-matching loss 
\begin{align}
L(f_{\theta},Y,H)= \| H-f_{\theta}(\tilde{H}_i) \|^2.
\end{align}
While our training setup shares architectural similarities with the supervised masked autoencoder from \cite{prior_inpainting}, the underlying inference paradigm is fundamentally different: rather than directly regressing the missing entries, our method learns a score function that enables iterative, likelihood-guided refinement, yielding significantly improved robustness to masking patterns and deployment-time distortions.

For the DDPM formulation, the noise schedule is governed by a monotonically increasing sequence $\{\beta_i\}_{i=1}^L$. Defining $\alpha_i = 1 - \beta_i$ and $\bar{\alpha}_i = \prod_{k \leq i} \alpha_k$, we simulate the forward diffusion process by sampling a timestep $i \sim \mathcal{U}[1, L]$, drawing Gaussian noise $z \sim \mathcal{N}(0, I)$, and constructing the noisy masked input as $\tilde{H}_i=\mathcal{A}\odot(\sqrt{\bar{\alpha}_i}H+\sqrt{1-\bar{\alpha}_i} z)$, where $\mathcal{A}$ denotes the masking pattern and $H$ is the clean channel matrix.

\begin{algorithm}
\caption{Diffusion Inference with Score Network} \label{alg1}
\begin{algorithmic}[1]
\STATE \textbf{Inputs:} Masked and noisy channel observations $Y$; pretrained score model $f_\theta$; observation noise variance $\sigma_{\text{obs}}^2$; noise levels $\{\sigma_{z_i}\}_{i=1}^L$; inner loop steps $M$; hyperparameters: $\alpha_0$, $\beta$, $\zeta$, $r_0$, $\beta_0$; SRS masking matrix $\mathcal{A}$
\STATE \textbf{Initialize:} $H_{\text{est}} \sim \mathcal{CN}(0, I)$
\FOR{each diffusion level $i = 1$ to $L$}
    \STATE Set noise level: $\sigma \leftarrow \sigma_{z_i}$
    \STATE Set step scaling: $r_i \leftarrow 1$ if $L < \zeta$, else $r_0^{L - \zeta}$
    \STATE Compute step size: $\alpha \leftarrow \alpha_0 r_i \cdot \frac{\sigma_{z_i}}{\sigma_L}$
    \FOR{each iteration $m = 1$ to $M$}
        \STATE Sample $z \sim \mathcal{N}(0, I)$
        \STATE Estimate score:
        $s(H_{\text{est}}) \leftarrow -\frac{H_{\text{est}} - f_\theta(\mathcal{A} \odot H_{\text{est}})}{\sigma}$
        \STATE Langevin update:
        \[
        H_{\text{est}} \leftarrow H_{\text{est}} + \alpha \left( s(H_{\text{est}}) + \frac{Y - \mathcal{A} \odot H_{\text{est}}}{\sigma_{\text{obs}}^2 + \sigma^2} \right) + \sqrt{2 \alpha \beta} \cdot z
        \]
    \ENDFOR
\ENDFOR
\STATE \textbf{Output:} In-painted channel estimate $H_{\text{est}}$
\end{algorithmic}
\end{algorithm}

\begin{algorithm}
\caption{Diffusion Inference with DDPM (VP-SDE)} \label{alg2}
\begin{algorithmic}[1]
\STATE \textbf{Inputs:} Masked and noisy channel observations $Y$; pretrained DDPM model $f_\theta$; DDPM noise schedule $\{\alpha_i, \beta_i\}_{i=1}^{L}$ with $\bar{\alpha}_i = \prod_{k \leq i} \alpha_k$; observation noise variance $\sigma_{\text{obs}}^2$; SRS masking matrix $\mathcal{A}$

\STATE \textbf{Initialize:} $H_{\text{est},L} \sim \mathcal{N}(0, I)$

\FOR{each reverse diffusion step $i = L$ down to $1$}
    \STATE Estimate score:
    $
    s(H_{\text{est},i}) \leftarrow \frac{\sqrt{\bar{\alpha}_i}}{1 - \bar{\alpha}_i} f_\theta(\mathcal{A} \odot H_{\text{est},i})
    $
    \STATE Estimate clean image:
    \[
    \hat{H}_0 \leftarrow \frac{1}{\sqrt{\bar{\alpha}_i}} \left( H_{\text{est},i} + (1 - \bar{\alpha}_i) \cdot s(H_{\text{est},i}) \right)
    \]
    \STATE Sample noise: $z \sim \mathcal{N}(0, I)$
    \STATE Compute posterior sample:
    \[
    H' \leftarrow \frac{\sqrt{\alpha_i}(1 - \bar{\alpha}_{i-1})}{1 - \bar{\alpha}_i} H_{\text{est},i} + \frac{\sqrt{\bar{\alpha}_{i-1}} \beta_i}{1 - \bar{\alpha}_i} \hat{H}_0 + \sqrt{\beta_i} \cdot z
    \]
    \STATE Gradient correction for likelihood:
    \[
    H_{\text{est},i-1} \leftarrow H' - \zeta \cdot \nabla_H \left( \frac{\|Y - \mathcal{A} \odot \hat{H}_0\|^2}{\sigma_{\text{obs}}^2} \right)
    \]
\ENDFOR

\STATE \textbf{Output:} Inpainted channel estimate $\hat{H}_0$
\end{algorithmic}
\end{algorithm}

The inference procedures for both score-based and DDPM-based inpainting are summarized in Algorithms~\ref{alg1} and \ref{alg2}, respectively. In both cases, the gradient of the log-likelihood term $\nabla_H \log p(Y \mid H)$ is computed under the assumption of additive white Gaussian noise. This term encapsulates the system-specific knowledge, such as the masking pattern, noise statistics, and antenna configuration, and plays a critical role in guiding the generation process toward physically plausible reconstructions. Importantly, the key distinction between our approach and conventional score-based or DPS-based samplers is that our model is explicitly trained to both denoise and inpaint from masked observations. As a result, its outputs directly estimate the underlying score function under the applied masking, enabling more accurate and robust inference in challenging SRS configurations.

\subsection{Model Architecture}

We adopt a masked autoencoder architecture with a Vision Transformer (ViT) backbone \cite{he2022masked}. The input is a two-channel image, where channels 0 and 1 correspond to the real and imaginary parts of the channel matrix. A 2D convolutional patch embedding layer with kernel size equal to stride partitions the input into non-overlapping patches, each mapped to a vector of fixed embedding dimension. After adding positional encodings, masked patches are removed and a CLS token is prepended. The resulting sequence is processed by lightweight ViT encoder blocks with reduced depth and fewer attention heads.
The encoder output is linearly projected to the decoder's embedding size. Masked positions are filled with a learnable token, positional encodings are added, and the sequence is passed through ViT decoder blocks. A final linear projection maps the output to the original image dimensions, followed by reshaping to reconstruct the inpainted channel.

\section{Experimental Evaluation}
\label{sec:experiments}

We evaluate the proposed scheme using Cluster Delay Line (CDL) models A, B, and C from \cite{3gpp38901}, assuming 100 MHz bandwidth, 30 kHz subcarrier spacing, and a 64-antenna array. The model input is a $2 \times 816 \times 64$ tensor representing real and imaginary parts across 816 subcarriers. We train with a 75\% masking ratio, enabling multiplexing of four users with non-overlapping subcarrier bands. Transformer depth is set to 4 for both encoder and decoder, with 8 attention heads; embedding sizes are 128 and 512, respectively, and patch size is $17 \times 4$. For the score-based method, we set $L = 3000$, $\sigma_{z_1} = 156.6$, $\sigma_{z_{t+1}}/\sigma_{z_t} = 0.995$, and inner loop count $M = 3$. For the DDPM-based method, we use $L = 1000$, $\beta_1 = 10^{-4}$ and $\beta_{1000} = 0.02$. Both models are trained for 800 epochs using Adam (learning rate = 0.001, batch size = 128) on 21,600 CDL examples (7,200 per type). Evaluation is performed on 7,200 held-out examples.

\begin{figure}[htbp]
  \centering
  \begin{subfigure}[b]{0.75\linewidth}
    \includegraphics[width=\textwidth]{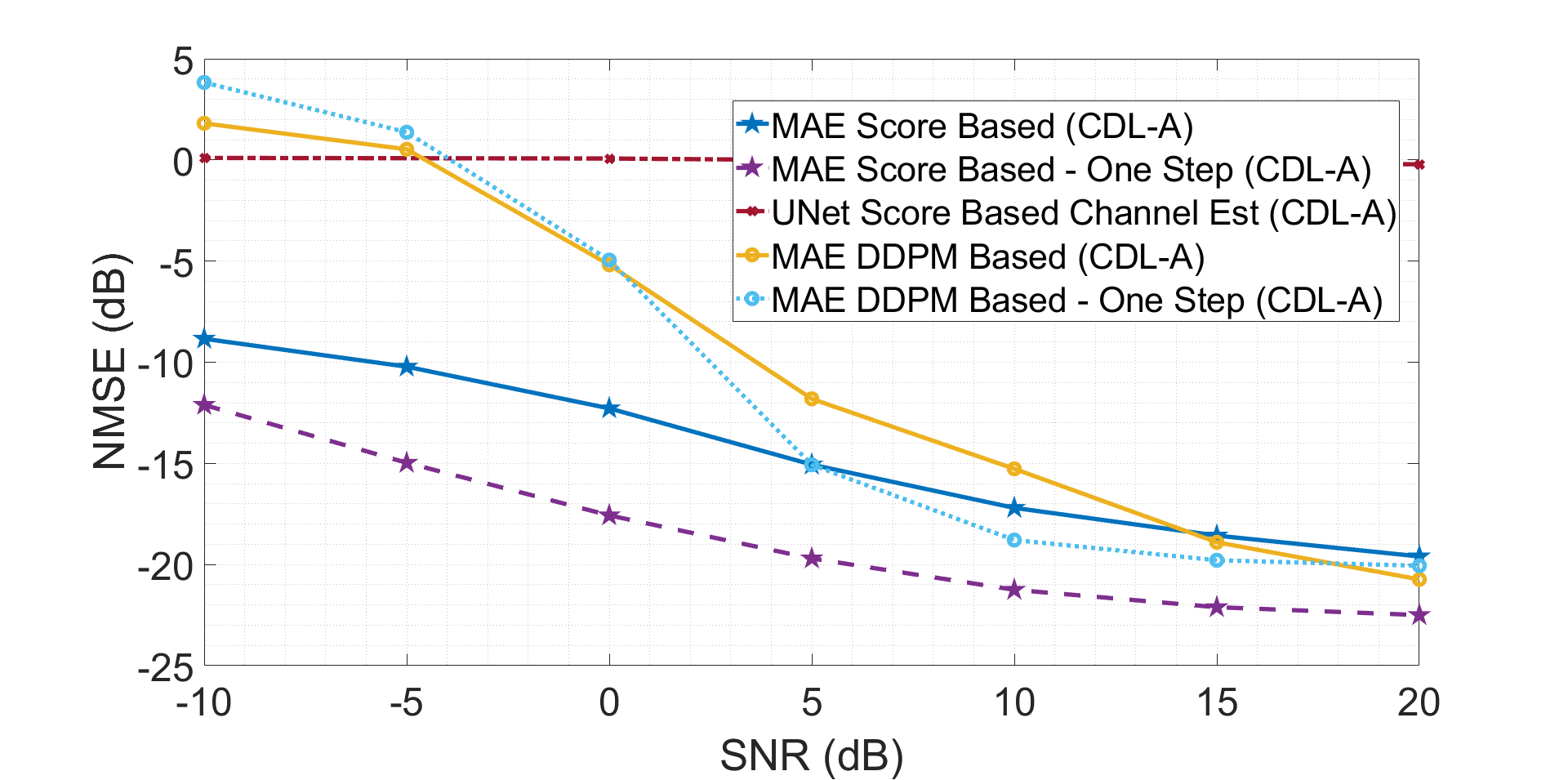}
    \caption{\small CDL-A: Proposed score-based method outperforms UNet and DDPM, especially at mid-to-high SNR.}
    \label{fig:subfig_a}
  \end{subfigure}
  \hfill
  \begin{subfigure}[b]{0.75\linewidth}
    \includegraphics[width=\textwidth]{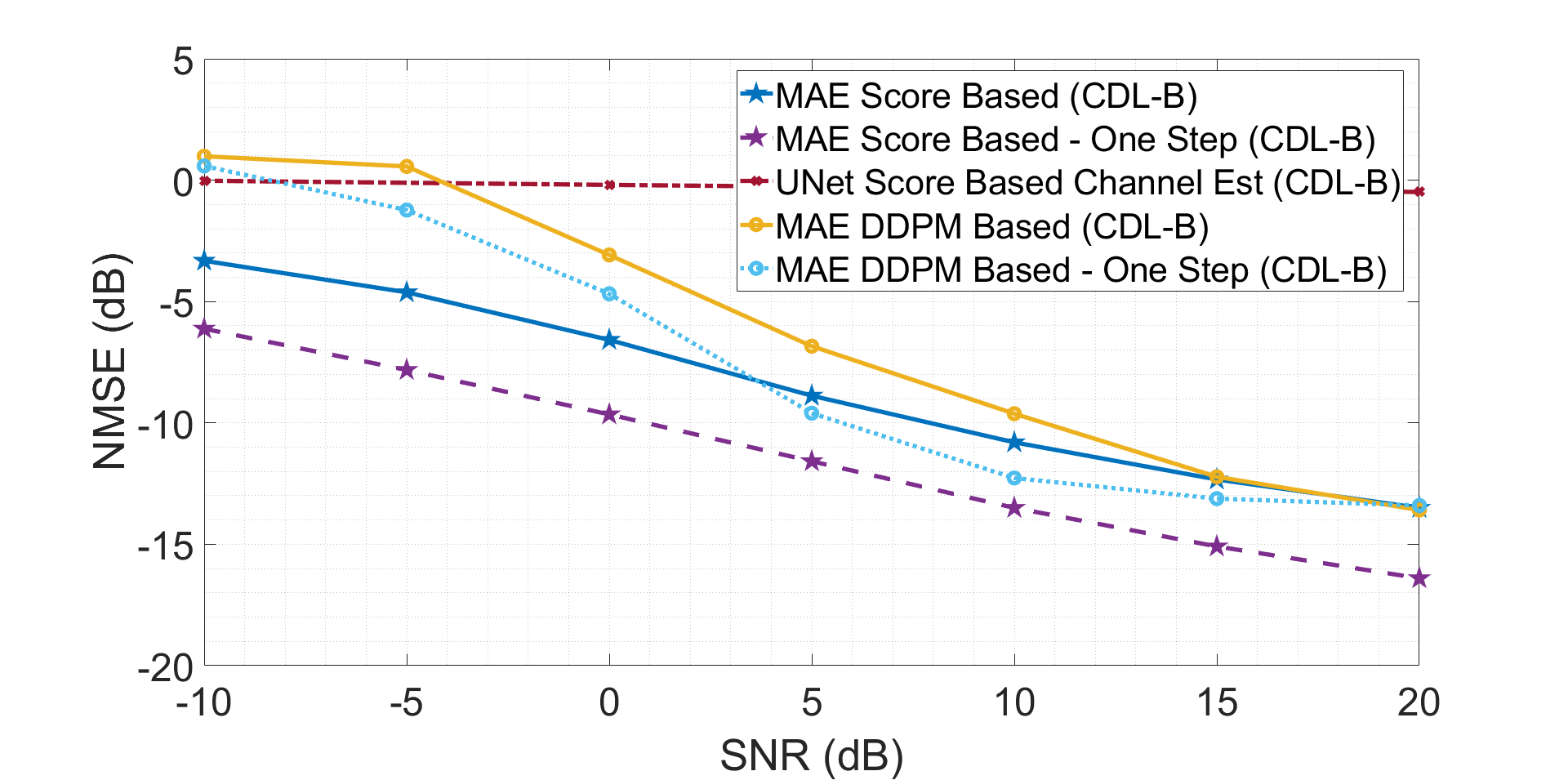}
    \caption{\small CDL-B: Score-based model maintains advantage over baselines across SNRs.}
    \label{fig:subfig_b}
  \end{subfigure}
  \hfill
  \begin{subfigure}[b]{0.75\linewidth}
    \includegraphics[width=\textwidth]{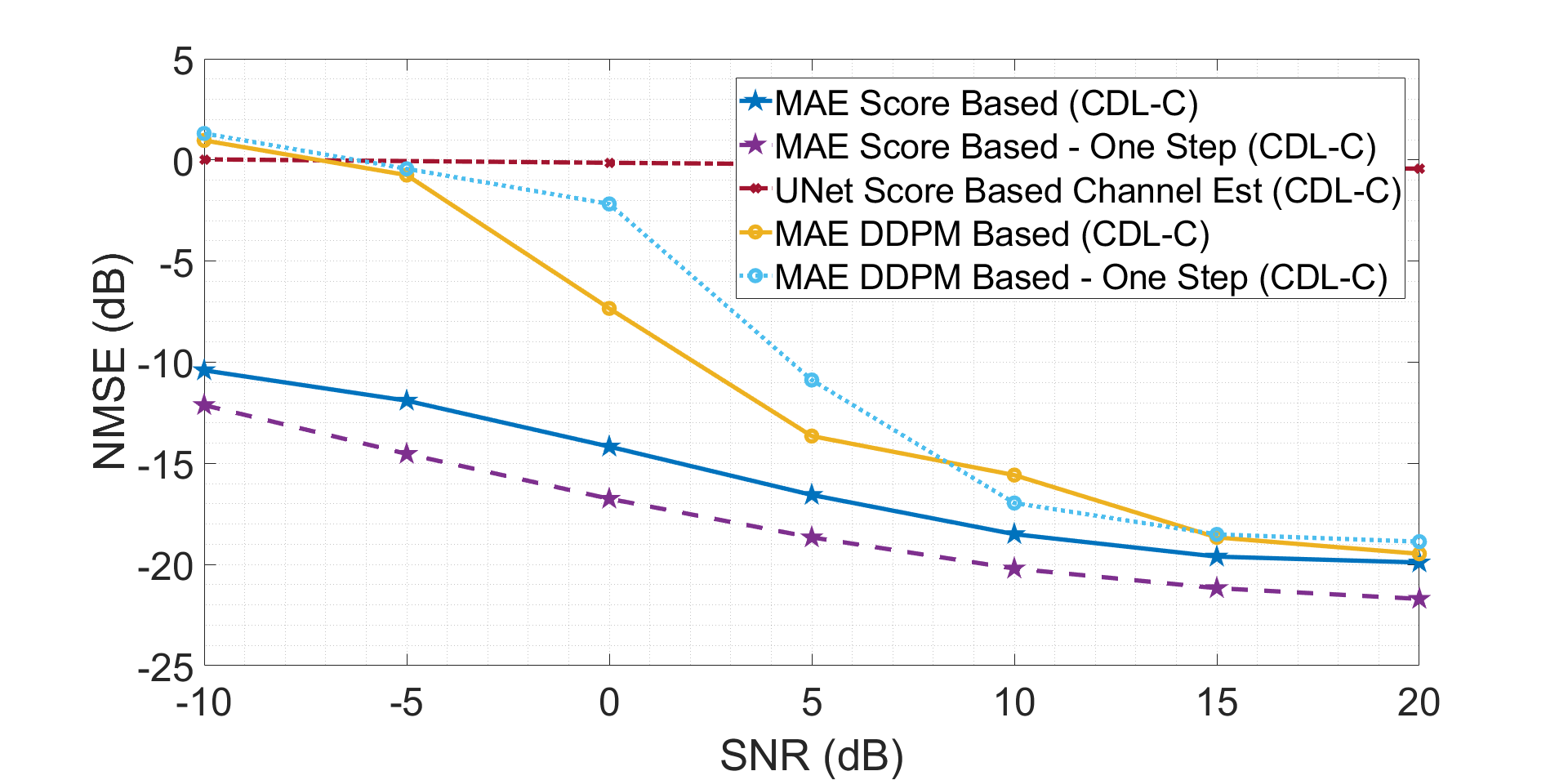}
    \caption{\small CDL-C: Proposed scheme shows largest margin over UNet at low SNR.}
    \label{fig:subfig_c}
  \end{subfigure}
  \caption{NMSE vs. SNR for CDL-A/B/C channels. Metrics are computed over sounded and reconstructed subcarriers.}
  \label{fig:results_1}
\end{figure}

As baselines, we include: (i) a score-based channel estimation model with a UNet backbone adapted from \cite{MIMO_Score_Est}, and (ii) a one-step version of our score-based model (i.e., inference without the inner loop), which matches our prior work \cite{prior_inpainting}. Fig.~\ref{fig:results_1} shows NMSE versus SNR for all methods. Across CDL-A, B, and C, both score-based and DDPM variants outperform the UNet baseline. Among the two, the score-based approach consistently yields better performance, particularly at higher SNRs. Notably, the one-step variant achieves the lowest NMSE overall -- highlighting that, under clean conditions, a single forward pass of the trained model (as in \cite{prior_inpainting}) can suffice.

The following experiments highlight the robustness of the score-based MAE model under distribution shift, where the test-time system model differs from that seen during training. This robustness stems from posterior-guided inference, wherein the model leverages a log-likelihood gradient term to align predictions with the true observation model. We evaluate robustness across five categories of test-time perturbations: additional masking, user interference, spatially varying SNR, non-Gaussian noise, and signal clipping.

\begin{figure}[htbp]
  \centering
  \begin{subfigure}[b]{0.75\linewidth}
    \includegraphics[width=\textwidth]{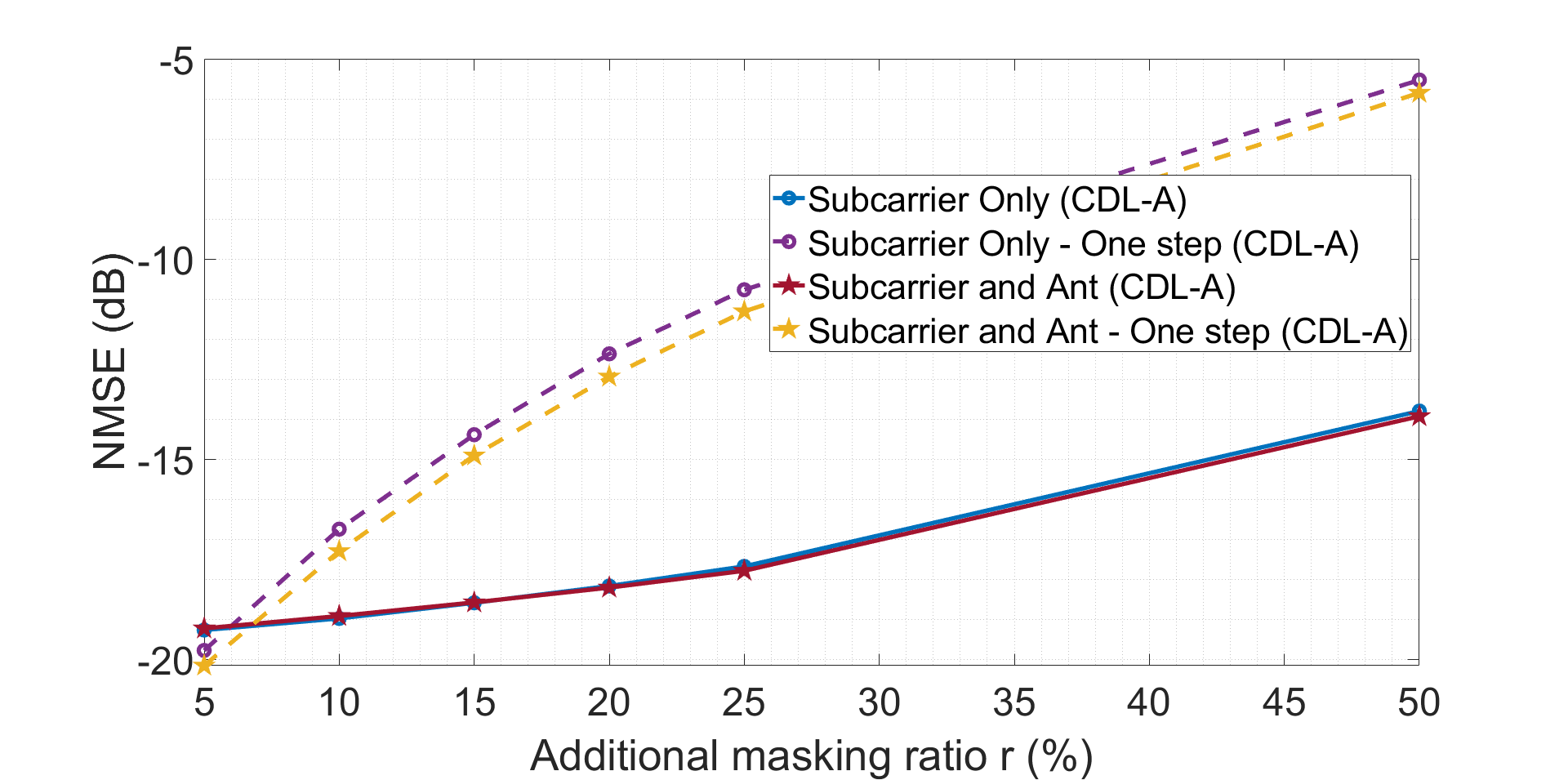}
    \caption{\small CDL-A: Diffusion model maintains low NMSE despite increased masking.}
    \label{fig:subfig_2a}
  \end{subfigure}
  \hfill
  \begin{subfigure}[b]{0.75\linewidth}
    \includegraphics[width=\textwidth]{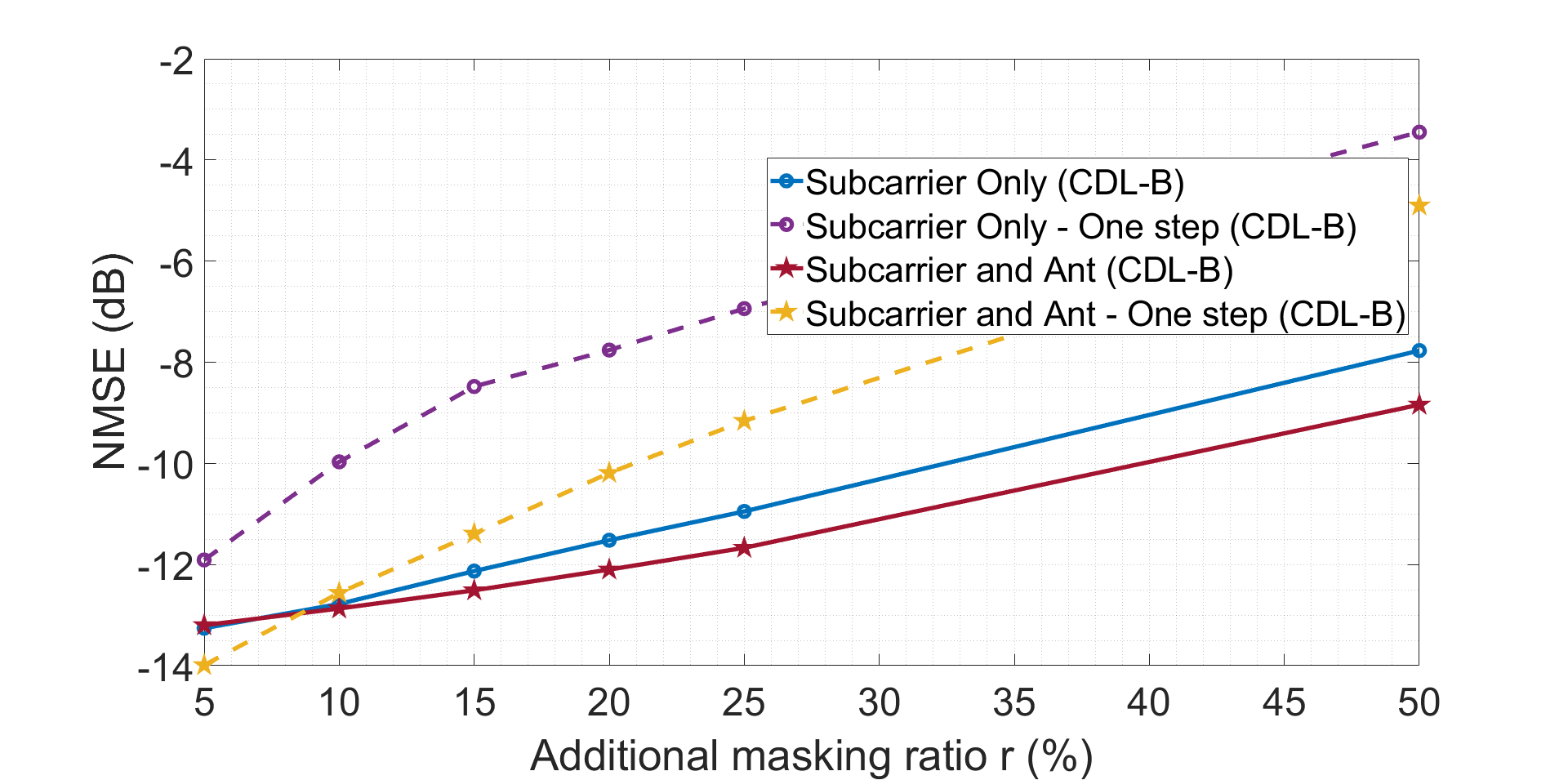}
    \caption{\small CDL-B: One-step baseline suffers sharp degradation under joint masking.}
    \label{fig:subfig_2b}
  \end{subfigure}
  \hfill
  \begin{subfigure}[b]{0.75\linewidth}
    \includegraphics[width=\textwidth]{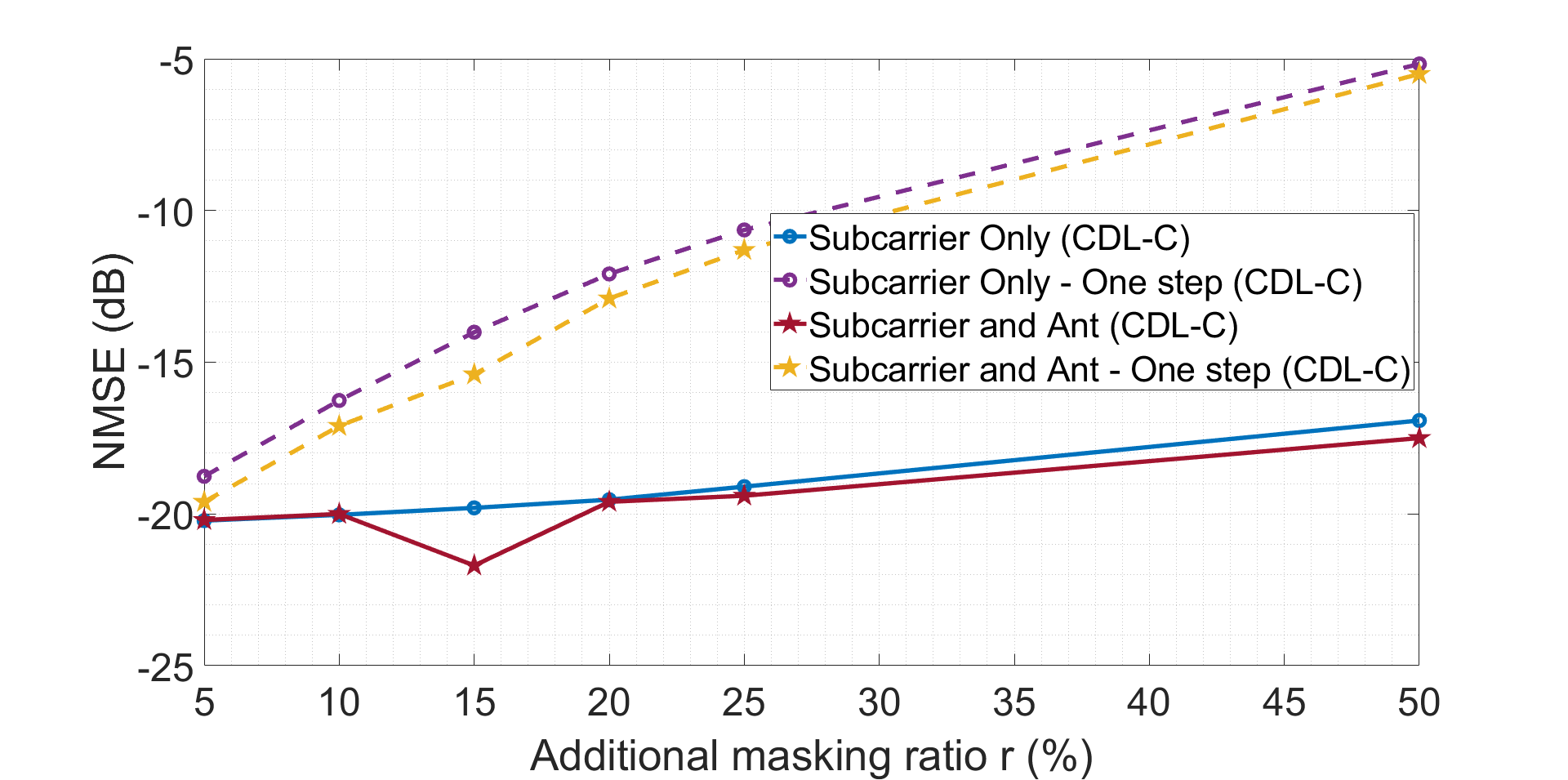}
    \caption{\small CDL-C: Diffusion approach sustains performance across masking types.}
    \label{fig:subfig_2c}
  \end{subfigure}
  \caption{NMSE vs. additional masking ratio $r$ (\%) at 20 dB SNR, with extra masking applied along subcarriers or both subcarriers and antennas.}
  \label{fig:results_2}
\end{figure}

\begin{figure}[h]
\centering
\includegraphics[width=0.75\columnwidth]{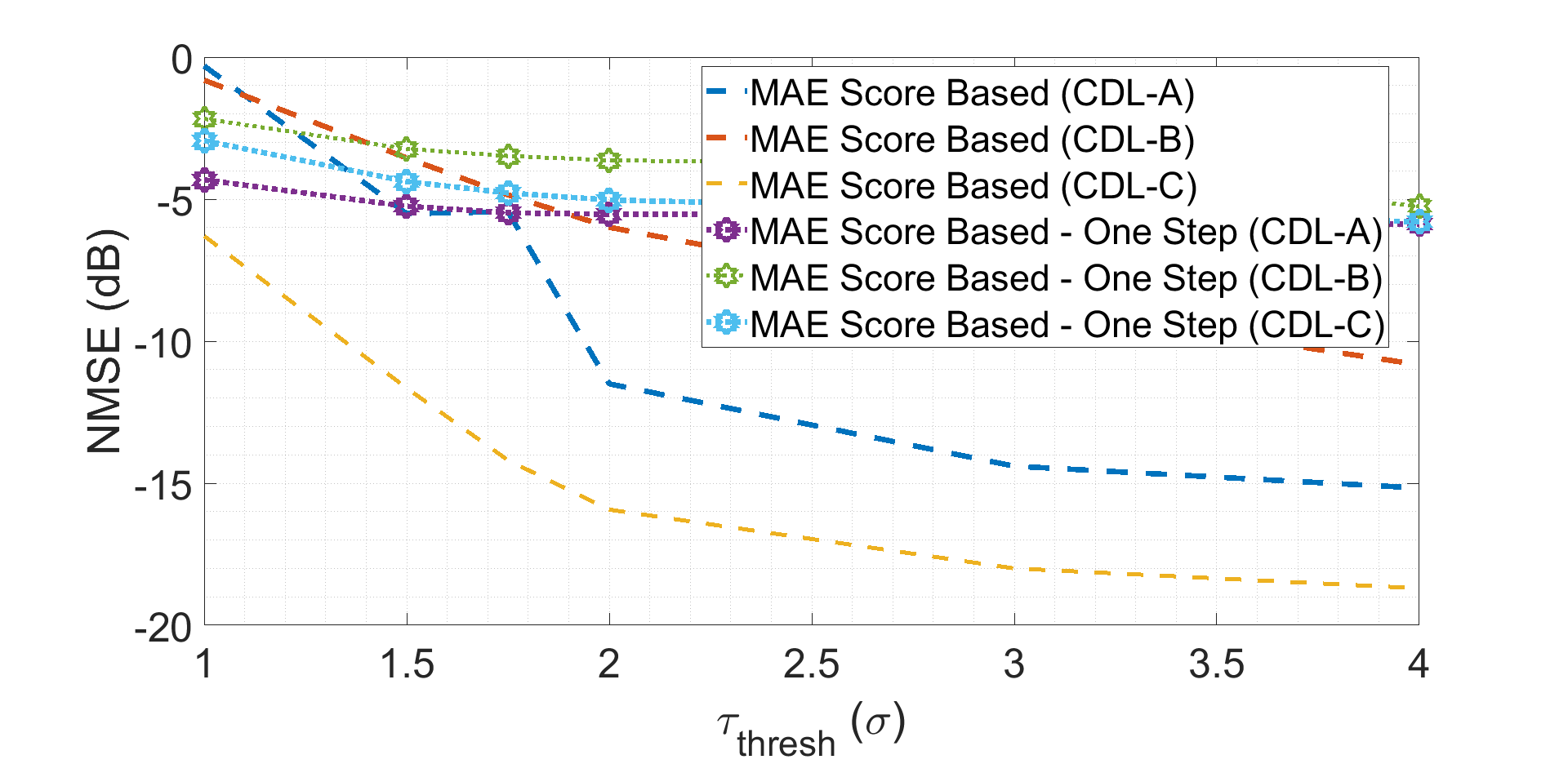}
\caption{Effect of clipping on NMSE at 20 dB SNR vs. clipping threshold (in standard deviation units).}
\label{fig:results_3}
\end{figure}

\begin{itemize}

\item \textbf{Additional Masking:} Superimposing on the structured masking pattern used during training, we randomly mask an additional $r\%$ of the pixels. Two settings are considered: masking only along subcarriers, and masking across both subcarriers and antennas. The evaluation SNR is fixed at 20~dB. As shown in Fig.~\ref{fig:results_2}, one-step inference suffers rapid NMSE degradation with increasing $r$, while the proposed method remains stable even with 50\% additional masking.

\item \textbf{User Interference on Sounded REs:} We simulate interference by adding pilot signals from two users: $Y = P_1 \odot H_1 + P_2 \odot H_2 + z_{\text{obs}}$, with $\text{SIR}=0$~dB and $\text{SNR}=20$~dB. One-step inference is run separately for $P_1$ and $P_2$. In contrast, the proposed scheme jointly estimates $H_{\text{est},1}$ and $H_{\text{est},2}$ using log-likelihood gradients  
\[
\frac{Y - \mathcal{A} \odot (P_1 \odot H_{\text{est},1} + P_2 \odot H_{\text{est},2})}{\sigma_{\text{obs}}^2 + \sigma^2}.
\] 
As shown in Table~\ref{tab1}, the joint diffusion-based approach achieves significantly better NMSE across all CDL types.

\item \textbf{Per-Pixel SNR Variation:} We sample the SNR for each pixel independently from $\mathcal{U}[-10, 20]$ to mimic spatial SNR heterogeneity. Results in Table~\ref{tab1} show that the proposed method maintains a performance edge over one-step inference for CDL-B and CDL-C.

\item \textbf{Non-Gaussian Observation Noise:} Here, the real and imaginary noise components are Laplace-distributed, tuned to match an average SNR of 20~dB. The log-likelihood gradient becomes  
$
\mathcal{A} \odot \frac{\text{sign}(y - \mathcal{A} \odot H_{est})}{\sqrt{(\sigma_{obs}^2 + \sigma^2)/2}}.
$
Under 50\% additional masking, the NMSE gap between diffusion-based inference and the baseline widens, as detailed in Table~\ref{tab1}.

\item \textbf{Signal Clipping:}  
We evaluate the model on the clipped system $Y = \mathcal{A} \odot \text{clip}(H, \tau_{\text{thresh}}) + z_{obs}$, which imposes nonlinear distortion. As shown in Fig.~\ref{fig:results_3}, the proposed approach maintains its advantage over the one-step baseline even under severe clipping and additional masking.

\end{itemize}

% \begin{figure}[h]
%     \includegraphics[width=\columnwidth]{Results_2.png}
%     \caption{NMSE against ground truth over sounded and reconstructed subcarrier indices vs additional uniform masking at an SNR of 20 dB}
%     \label{fig:results_2}
% \end{figure}

\begin{table}[h]
\centering
\resizebox{0.75\columnwidth}{!}{

    \begin{tabular}{|l|c|c|} % | for vertical lines, l/c/r for left/center/right alignment
        \hline % Horizontal line
         Experiment & Diffusion & One-step Baseline \\ % Column headers
        \hline % Horizontal line
        Joint Inpainting, CDL-A & \textbf{-7.69} & -0.16 \\
        Joint Inpainting, CDL-B & \textbf{-5.56} & -0.02 \\
        Joint Inpainting, CDL-C & \textbf{-13.67} & -0.15  \\
        Per-RE SNR Variation, CDL-A & -16.38 & \textbf{-16.48} \\
        Per-RE SNR Variation, CDL-B & \textbf{-9.40} & -8.88 \\
        Per-RE SNR Variation, CDL-C &  \textbf{-16.76} & -15.77 \\
        Laplace Observation Noise, CDL-A & \textbf{-19.80} & -5.85 \\
        Laplace Observation Noise, CDL-B & \textbf{-13.22} & -4.92 \\
        Laplace Observation Noise, CDL-C & \textbf{-20.05} & -5.47 \\
        \hline        
    \end{tabular}
}
\caption{\small NMSE (dB) across distortion scenarios unseen during training. The proposed diffusion-based scheme consistently outperforms one-step inference, especially under joint inpainting and non-Gaussian noise. Bold indicates lower (better) NMSE. \normalsize}
\normalsize
\label{tab1}
\end{table}

For completeness, we evaluated two unsupervised baselines: a masked autoencoder trained with reconstruction loss, and a variant trained with an adversarial loss using a ViT discriminator. At 20dB SNR, these yielded NMSEs of –6.17dB and –5.22dB, respectively, substantially worse than our proposed methods. We therefore omit them from further analysis.

\textit{Limitations:} While training complexity remains comparable to \cite{prior_inpainting}, inference with diffusion models is several orders of magnitude more expensive than one-step prediction. This work serves as a proof of concept; future exploration may leverage faster generative techniques such as DDIM \cite{SongME21} or flow matching \cite{LipmanBCN22} to reduce inference cost.

\vspace{-0.05in}

\section{Conclusion}

\vspace{-0.05in}

We proposed a diffusion-based framework for robust SRS channel inpainting in 5G systems, capable of adapting to deployment-time impairments such as mask mismatch, interference, and non-Gaussian noise. Unlike prior one-shot reconstruction approaches, our method leverages posterior-guided inference to improve generalization under distribution shift. While the iterative nature of diffusion imposes higher inference cost, our formulation is compatible with fast variants such as DDIM and flow matching. These results position diffusion-guided inpainting as a promising tool for scalable, high-coverage CSI acquisition in future wireless networks.

% if have a single appendix:
%\appendix[Proof of the Zonklar Equations]
% or
%\appendix  % for no appendix heading
% do not use \section anymore after \appendix, only \section*
% is possibly needed

% use appendices with more than one appendix
% then use \section to start each appendix
% you must declare a \section before using any
% \subsection or using \label (\appendices by itself
% starts a section numbered zero.)
%

% \appendices
% \section{Proof of the First Zonklar Equation}
% Appendix one text goes here.

% % you can choose not to have a title for an appendix
% % if you want by leaving the argument blank
% \section{}
% Appendix two text goes here.

% % use section* for acknowledgment
% \section*{Acknowledgment}

% The authors would like to thank...

% Can use something like this to put references on a page
% by themselves when using endfloat and the captionsoff option.
\if CLASSOPTIONcaptionsoff
  \newpage
\fi

% trigger a \newpage just before the given reference
% number - used to balance the columns on the last page
% adjust value as needed - may need to be readjusted if
% the document is modified later
%\IEEEtriggeratref{8}
% The "triggered" command can be changed if desired:
%\IEEEtriggercmd{\enlargethispage{-5in}}

% references section

% can use a bibliography generated by BibTeX as a .bbl file
% BibTeX documentation can be easily obtained at:
% http://mirror.ctan.org/biblio/bibtex/contrib/doc/
% The IEEEtran BibTeX style support page is at:
% http://www.michaelshell.org/tex/ieeetran/bibtex/
%\bibliographystyle{IEEEtran}
% argument is your BibTeX string definitions and bibliography database(s)
%\bibliography{IEEEabrv,../bib/paper}
%
% <OR> manually copy in the resultant .bbl file
% set second argument of \begin to the number of references
% (used to reserve space for the reference number labels box)
\bibliographystyle{plain}
\bibliography{refs_new}

\begin{thebibliography}{10}

\bibitem{3gpp38901}
{3rd Generation Partnership Project (3GPP)}.
\newblock {Study on channel model for frequencies from 0.5 to 100 GHz (Release 16)}.
\newblock Technical Report TR 38.901 V16.1.0, 3GPP, December 2019.
\newblock Available: \url{https://www.3gpp.org/ftp/Specs/archive/38_series/38.901/}.

\bibitem{ADPS2}
Asad Aali, Giannis Daras, Brett Levac, Sidharth Kumar, Alex Dimakis, and Jon Tamir.
\newblock Ambient diffusion posterior sampling: Solving inverse problems with diffusion models trained on corrupted data.
\newblock In {\em The Thirteenth International Conference on Learning Representations}, 2025.

\bibitem{prior_inpainting}
Usman Akram, Fan Zhang, Shawn Ma, Yang Li, and Haris Vikalo.
\newblock Super capacity srs design for 5g and beyond using channel in-painting.
\newblock In {\em ICASSP 2025 - 2025 IEEE International Conference on Acoustics, Speech and Signal Processing (ICASSP)}, pages 1--5, 2025.

\bibitem{MIMO_Score_Est}
Marius Arvinte and Jonathan~I Tamir.
\newblock Mimo channel estimation using score-based generative models.
\newblock {\em IEEE Transactions on Wireless Communications}, 22(6):3698--3713, 2022.

\bibitem{DPS}
Hyungjin Chung, Jeongsol Kim, Michael~Thompson Mccann, Marc~Louis Klasky, and Jong~Chul Ye.
\newblock Diffusion posterior sampling for general noisy inverse problems.
\newblock In {\em The Eleventh International Conference on Learning Representations}, 2023.

\bibitem{ADPS}
Giannis Daras, Kulin Shah, Yuval Dagan, Aravind Gollakota, Alex Dimakis, and Adam Klivans.
\newblock Ambient diffusion: Learning clean distributions from corrupted data.
\newblock {\em Advances in Neural Information Processing Systems}, 36:288--313, 2023.

\bibitem{trad_5}
Dian Fan, Feifei Gao, Yuanwei Liu, Yansha Deng, Gongpu Wang, Zhangdui Zhong, and Arumugam Nallanathan.
\newblock Angle domain channel estimation in hybrid millimeter wave massive mimo systems.
\newblock {\em IEEE Transactions on Wireless Communications}, 17(12):8165--8179, 2018.

\bibitem{trad_4}
Zhen Gao, Chen Hu, Linglong Dai, and Zhaocheng Wang.
\newblock Channel estimation for millimeter-wave massive mimo with hybrid precoding over frequency-selective fading channels.
\newblock {\em IEEE Communications Letters}, 20(6):1259--1262, 2016.

\bibitem{he2022masked}
Kaiming He, Xinlei Chen, Saining Xie, Yanghao Li, Piotr Doll{\'a}r, and Ross Girshick.
\newblock Masked autoencoders are scalable vision learners.
\newblock In {\em Proceedings of the IEEE/CVF conference on computer vision and pattern recognition}, pages 16000--16009, 2022.

\bibitem{holma2011lte}
Harri Holma and Antti Toskala.
\newblock {\em LTE for UMTS: Evolution to LTE-advanced}.
\newblock John Wiley \& Sons, 2011.

\bibitem{trad_2}
Amin Khansefid and Hlaing Minn.
\newblock On channel estimation for massive mimo with pilot contamination.
\newblock {\em IEEE Communications Letters}, 19(9):1660--1663, 2015.

\bibitem{Yang2023AiChAug}
Yang Li, Yeqing Hu, Kyungsik Min, HyoYol Park, Hayoung Yang, Tiexing Wang, Junmo Sung, Ji-Yun Seol, and Charlie~Jianzhong Zhang.
\newblock Artificial intelligence augmentation for channel state information in 5g and 6g.
\newblock {\em IEEE Wireless Communications}, 30(1):104--110, 2023.

\bibitem{LipmanBCN22}
Yaron Lipman, Ricky T.~Q. Chen, Heli Ben-Hamu, Maximilian Nickel, and Matthew Le.
\newblock Flow matching for generative modeling.
\newblock {\em CoRR}, abs/2210.02747, 2022.

\bibitem{trad_1}
Yantao Qiao, Songyu Yu, Pengcheng Su, and Lijun Zhang.
\newblock Research on an iterative algorithm of ls channel estimation in mimo ofdm systems.
\newblock {\em IEEE Transactions on Broadcasting}, 51(1):149--153, 2005.

\bibitem{SongME21}
Jiaming Song, Chenlin Meng, and Stefano Ermon.
\newblock Denoising diffusion implicit models.
\newblock In {\em International Conference on Learning Representations (ICLR)}, 2021.
\newblock arXiv:2010.02502.

\bibitem{trad_3}
You You and Li~Zhang.
\newblock Bayesian matching pursuit-based channel estimation for millimeter wave communication.
\newblock {\em IEEE Communications Letters}, 24(2):344--348, 2020.

\end{thebibliography}

\end{document}